\documentclass[runningheads]{llncs}
\usepackage[T1]{fontenc}
\usepackage{amsmath}
\usepackage{graphicx}
\usepackage[]{color-edits}
\addauthor{nz}{red}
\addauthor{cz}{blue}

\begin{document}
\title{On the Foundation  Model for Cardiac MRI Reconstruction}
%
%
\author{Chi Zhang\inst{1} \and
Michael Loecher \inst{1} \and
Cagan Alkan \inst{1} \and
Mahmut Yurt \inst{1} \and
Shreyas S. Vasanawala\inst{1} \and
Daniel B. Ennis\inst{1}\orcidID{0000-0001-7435-1311}}
\authorrunning{F. Author et al.}
%
\institute{Stanford University, Stanford, CA, USA}
\maketitle              
\begin{abstract}
In recent years, machine learning (ML) based reconstruction has been widely investigated and employed in cardiac magnetic resonance (CMR) imaging. ML-based reconstructions can deliver clinically acceptable image quality under substantially accelerated scans. ML-based reconstruction, however, also requires substantial data and computational time to train the neural network, which is often optimized for a fixed acceleration rate or image contrast. In practice, imaging parameters are often tuned to best suit the diagnosis, which may differ from the training data. This can result in degraded image quality, and multiple trained networks are needed to fulfill the clinical demands. In this study, we propose a foundation model that uses adaptive unrolling, channel-shifting, and Pattern and Contrast-Prompt-UNet (PCP-UNet) to tackle the problem. In particular, the undersampled data goes through a different number of unrolled iterations according to its acceleration rate. Channel-shifting improves reconstructed data quality. The PCP-UNet is equipped with an image contrast and sampling pattern prompt. \emph{In vivo} CMR experiments were performed using mixed combinations of image contrasts, acceleration rates, and (under)sampling patterns. The proposed foundation model has significantly improved image quality for a wide range of CMR protocols and outperforms the conventional ML-based method. 

\end{abstract}

\section{Introduction}
Cardiac magnetic resonance (CMR) imaging is a widely applied clinical tool for the diagnosis of various cardiac diseases. CMR exams are time-consuming and research currently focuses on ways to reduce exam times. Most approaches to faster CMR exams rely on undersampling the acquired \emph{k}-space data and using advanced reconstruction methods to generate images of acceptable clinical quality. In particular, recent advances in machine learning (ML) based reconstruction \cite{modl,dl2,dl3,dl4,raki,julioReview} have enabled substantial increases in acceleration rates (e.g. 10-20x) for CMR, which extends its availability to patients, as well as enabling advanced CMR techniques \cite{flow}. These approaches, however, are generally bespoke for each acquisition protocol. ML-based reconstruction approaches also require substantial amounts of data and compute time to be trained, but once trained provide very fast reconstruction times that outperform conventional methods (e.g. compressed sensing). 

Existing ML-based reconstructions can be split into two major categories: Scan-specific and data-driven approaches. In the former, neural networks learn to interpolate \emph{k}-space from the fully sampled \emph{k}-space center (i.e. calibration lines) of the image to be reconstructed \cite{raki,sraki,rraki}. In the latter, a neural network is trained using a large imaging database \cite{modl,dl2,dl3,dl4}. Restricted by the limited number of calibration lines, scan-specific approaches are typically forced to employ compact network designs, and perform poorly when reconstructing images acquired with high acceleration rates. In contrast, when sufficient data is available for training, the data-driven approaches leverage more advanced network designs, including transformers \cite{dltransformer} and diffusion models \cite{diffusion}. The data-driven approach can pushing the acceleration rates even higher while maintaining satisfactory image quality.  

To optimize image quality and minimize the risk of distribution shifts, protocol-specific learning is commonly adapted in data-driven approaches \cite{modl,dl2,flow}, which requires that the images for used training and inference are matched in image contrast, sampling patterns, and acceleration rates. Any mistmatch  between training and inference will typically lead to a significant degradation in image quality \cite{modl}. This is a major practical issue of the data-driven approaches because the imaging protocol and parameters are generally refined for each CMR exam. In a typical CMR exam, images of multiple views and contrast are acquired to assist diagnosis. Further adjustments in sampling pattern and acceleration rates may also be needed \cite{spiral,cine,phaseconstrast}. One could deploy multiple networks matched to all exam possibilities, but this adds substantial complexity and is not generally feasible. 

In this work, we propose a universal foundation model for machine learning-based CMR reconstruction that allows reconstruction of mixed contrast, views, sampling patterns, and acceleration rates. To do so we employ an unrolled network \cite{modl} architecture to solve a regularized reconstruction problem. We inherit the analytical conclusion from compressed sensing \cite{cs}, to handle different acceleration rates by using different numbers of unrolled steps. We adapt the Prompt-UNet \cite{helopipu} for mixed image contrasts and views, and extend it with additional sampling embedding, which we refereed as Pattern and Contrast-Prompt-UNet (PCP-UNet) to optimize image quality of different sampling patterns. To better reconstruct various sampling patterns, we also apply a novel channel-shift technique to achieve a wider receptive field in the PCP-UNet with minor computational overhead. 

Experiments were conducted using \emph{in vivo} CMR data from the CMRxRecon 2024 challenge. The proposed method outperformed a simple unrolled network with a fixed number of unrolled steps using either UNet or Prompt-UNet. The proposed PCP-UNet outperformed Prompt-UNet using fixed or adaptive unrolls. Statistical results showed improved structural similarity index measure (SSIM) were obtained using PCP-Unet, which is suitable for handling reconstruction of images with a variety of image contrast, views, sampling patterns, and acceleration rates.


\section{Methodology}

\subsection{Regularized Reconstruction and Unrolling} 

Let $\bf E_{\Omega}$ be the forward acquisition model associated with sampling pattern $\bf \Omega$, $\bf y$ denotes the acquired \emph{k}-space data, and the regularization term is $\mathcal{R}(\mathbf{x})$. The regularized reconstruction problem is modeled as: 




\begin{align}\label{eq:1}
\arg\min_{\mathbf{x}} \| {\bf E_{\Omega} x - y} \|_2^2 + \mathcal{R}(\mathbf{x})
\end{align}

A common approach to solve (\ref{eq:1}) is model unrolling via variable splitting \cite{cs,modl}, which introduces an auxiliary variable $\bf u$ to the original problem:

\begin{align}\label{eq:2}
\arg\min_{\mathbf{x}} & \| {\bf E_{\Omega} x - y} \|_2^2 + \mathcal{R}(\mathbf{u}) \\
& s.t. \ \ \ \ {\bf x = u} \nonumber
\end{align}

\noindent Eq. \ref{eq:2} can be solved by iteratively solving the sub-problems:



\begin{align}\label{eq:network}
\mathbf{u}^{(i)} &= \arg\min_{\mathbf{u}} \mu \| \mathbf{x}^{(i-1)} - \mathbf{u} \|_2^2 + \mathcal{R}(\mathbf{u})
\end{align}
\begin{align}\label{eq:dc}
\mathbf{x}^{(i)} &= \arg\min_{\mathbf{x}} \| \mathbf{E}_{\bf \Omega} \mathbf{x} - \mathbf{y} \|_2^2 + \mu \| \mathbf{u}^{(i)} - \mathbf{x} \|_2^2
\end{align}
where Eq. \ref{eq:network} is often solved using a neural network \cite{modl}. Eq. \ref{eq:dc} is the data-consistency (DC) term, which is solved via \emph{e.g.} gradient descent \cite{dl2,flow,helopipu} or the conjugate gradient method \cite{modl}.

\begin{figure}
\begin{center}
\includegraphics[width=0.8\textwidth]{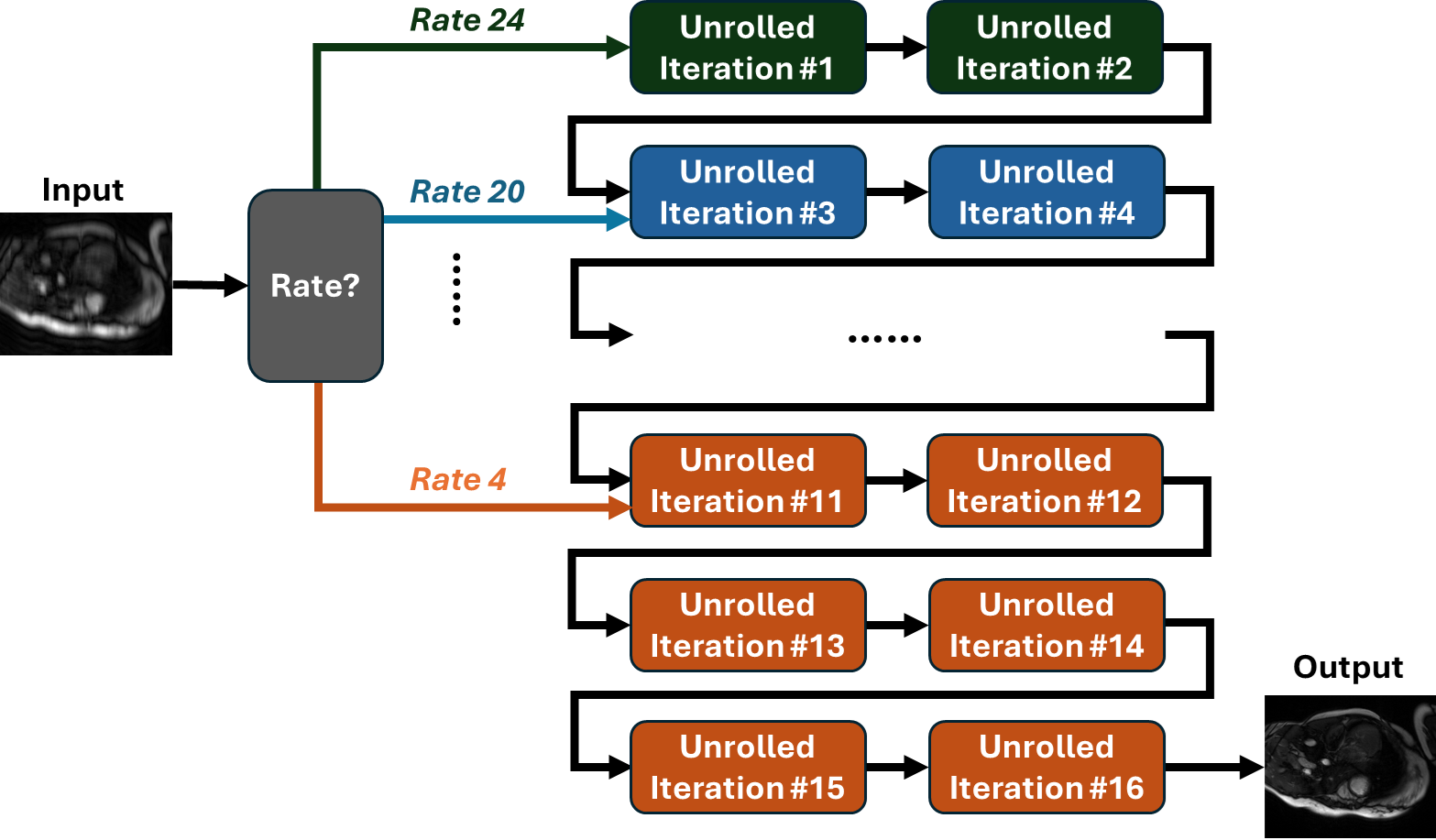}
\end{center}
\caption{Adaptive unrolling. Undersampled measurements are fed to a specific unrolled iteration according to the acceleration rate. Images of higher acceleration rate proceed through more unrolled iterations than those of lower acceleration rates. Each unrolled iteration has its own network regularizer and conjugate gradient with parameters independently determined relative to the other unrolled iterations.} \label{fig1}
\end{figure}

\subsection{Adaptive Unrolling}
The acceleration rate impacts the condition number of $\bf E_{\Omega}$. This suggests the number of unrolled iterations (UI) required for  convergence is different for different acceleration rates. This is also implied by previous works, where highly accelerated acquisitions were reconstructed using more UI to improve image quality \cite{helopipu}. To reconstruct images for a wide range of acceleration rates, we propose to use adaptive unrolling that performs fewer (more) UI for relatively low (high) acceleration rates. Fig.\ref{fig1} depicts the proposed unrolled network. Specifically, our unrolled cascade is implemented using up to sixteeen UI. Each UI has its own neural network and conjugate gradient solver, whose parameters are independent from the other UI. An input image enters the unrolled cascade from a specific UI according to its acceleration rate. As such, images with the highest acceleration rate go through all sixteen UI, and images with lower acceleration rates will be reconstructed via the latter UI only. As a result, each UI has its own neural network and conjugate gradient solver optimized for inputs with specific level of artifacts.

\begin{figure}

\begin{center}
\includegraphics[width=0.8\textwidth]{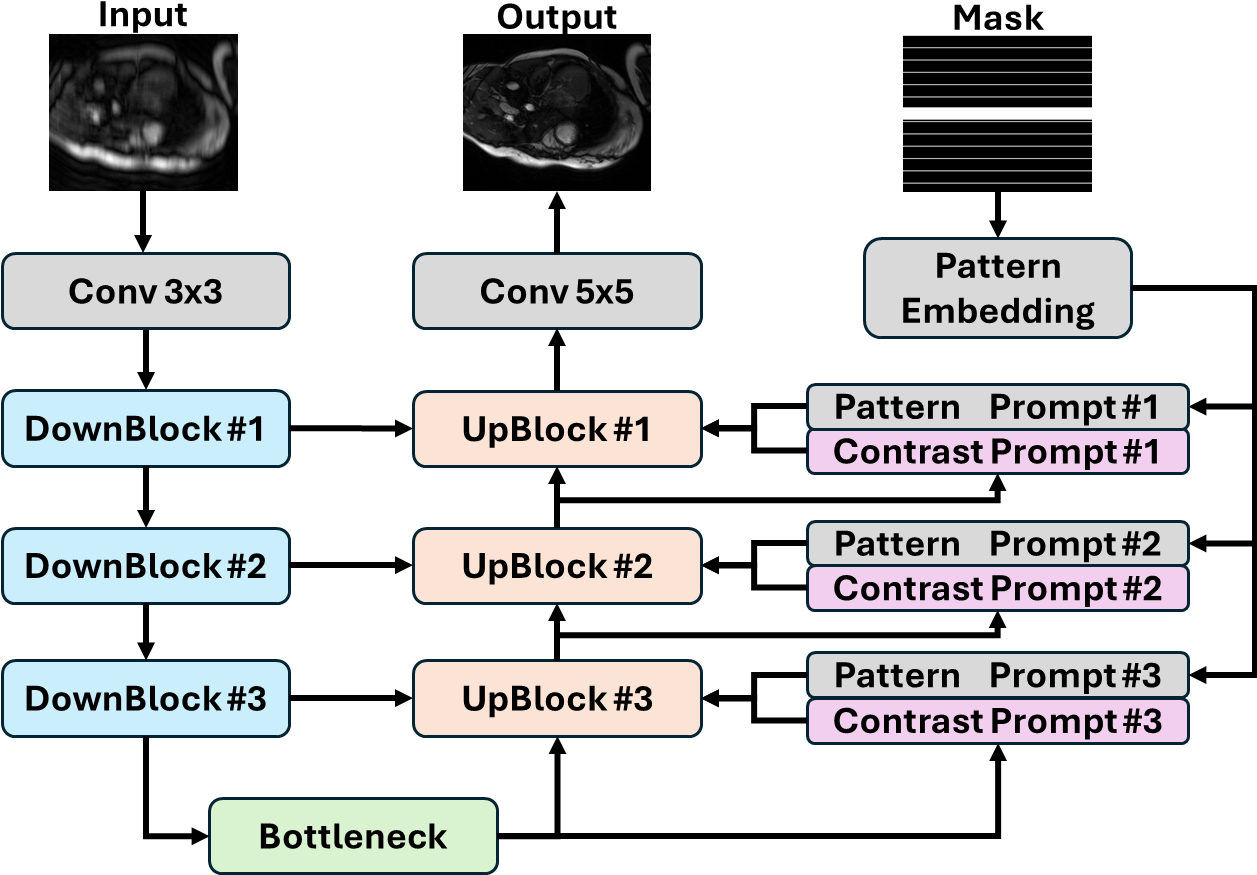}
\end{center}
\caption{Pattern and Contrast-Prompt (PCP) UNet. PCP-UNet inherits the Prompt-UNet concept with additional \emph{k}-space undersampling pattern prompt modules that are similar to the contrast prompt modules in Prompt-UNet.} \label{fig2}
\end{figure}

\subsection{Pattern and Contrast-Prompt (PCP) UNet}
Based on Prompt-UNet \cite{helopipu}, we extend the contrast prompt with a pattern-prompt to reconstruct images of mixed contrasts and sampling patterns (Fig.\ref{fig2}). The pattern prompt modules receive pattern embeddings as input, followed by attention and convolutions, which is defined in \cite{helopipu}. Feature maps from image domain convolution, contrast-prompt, and pattern-prompt are concatenated along the channel dimension. As such, the neural network step (Eq. \ref{eq:network}) is solved by a function of ${\bf x}^{(i)}$ and $\bf \Omega$ to obtain reconstructed images for different sampling patterns.

\begin{figure}
\includegraphics[width=\textwidth]{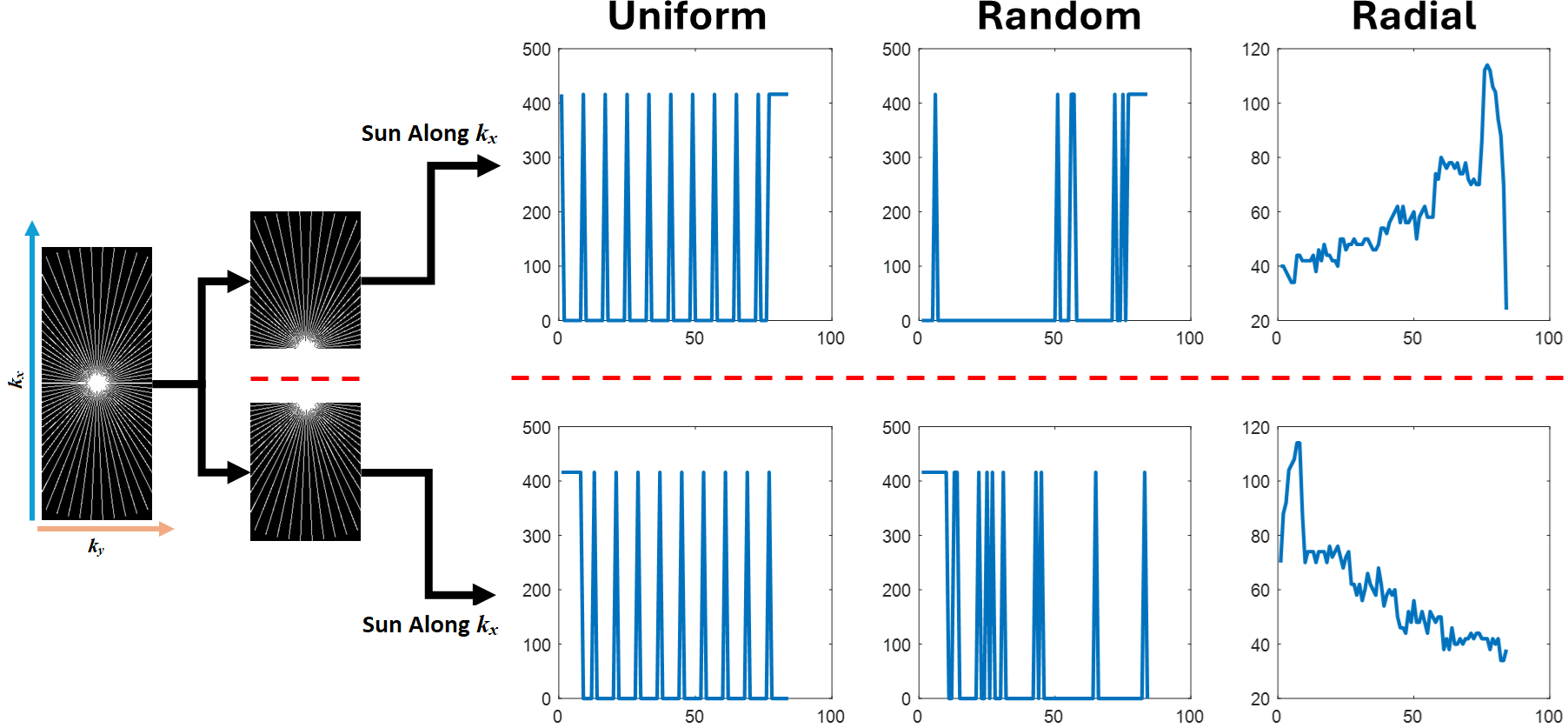}
\caption{\emph{k}-space sampling pattern embeddings were estimated from statistical information of the sampling mask, with an example of statistics along $k_x$ direction. The sampling mask is first split in two halves along the $k_x$ direction, in order to distinguish between uniform and random sampling. Summation along $k_x$ is applied to obtain the total number of samples as well as its distribution. The mean and variance of the sample number and the spacing between non-zero data points are computed as statistical feature of the sampling mask.} \label{fig3}
\end{figure}

In this study we generate \emph{k}-space sampling pattern embeddings via statistical features of the sampling mask (Fig \ref{fig3}), in order to produce numerical features distinguishable between three commonly used sampling patterns: uniform, Gaussian random, and simulated radial pattern on Cartesian grid. Specifically, we first split the sampling mask by half along the \emph{$k_x$} direction, which is intentionally aimed to differentiate Gaussian random patterns of varied sampling densities. Then we accumulate the total sample numbers along the phase and frequency encoding directions of each half of the sampling mask. This is important for distinguishing between uniform and random sampling. We further compute the statistics of the spacing between samples along each direction. These embedded features can be used to distinguish between sampling patterns and also reflects the sampling density of the sampling mask.

\begin{figure}
\includegraphics[width=\textwidth]{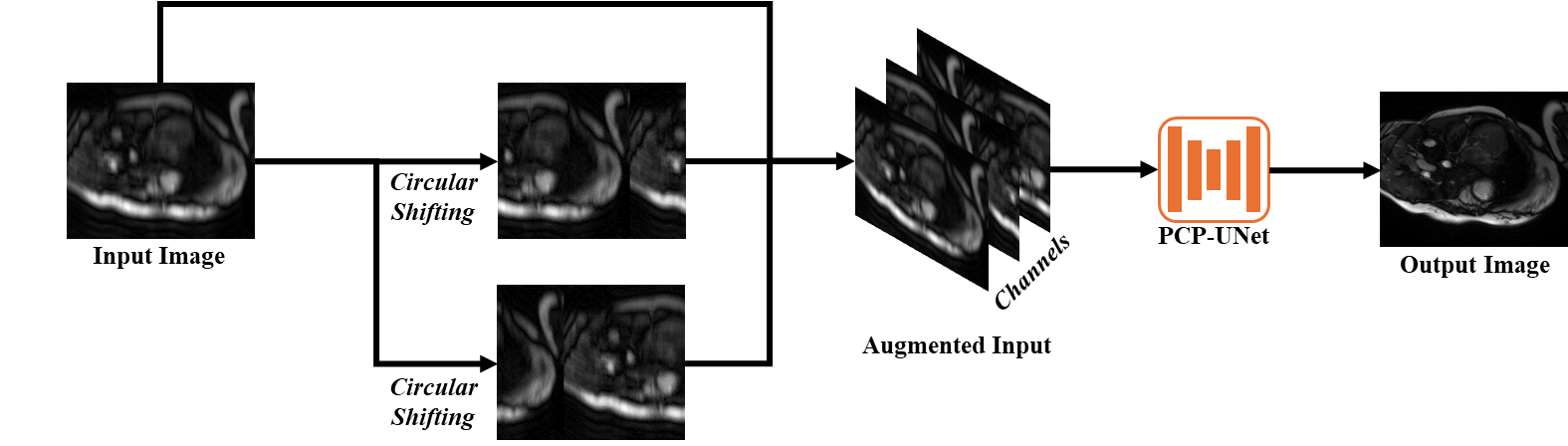}
\caption{Enlarging receptive field via channel-shifting. Circular shifting along the sub-sampled direction(s) is performed to produce shifted replicas of the input image. The replicas are subsequently concatenated with the input along channel dimension to produce an augmented input, which is fed into a regular CNN with additional channels in its input layer.} \label{figCSCNN}
\end{figure}

\subsection{Enlarging Receptive Field via Channel-Shift}
Various sampling patterns result in different point-spread functions (PSF) in the image domain. Notably, the PSF can span the entire FOV due to the presence of a fully sampled center, as well as unevenly distributed sampling positions. This indicates a global receptive field is necessary to better distinguish between PSFs of different sampling patterns, which can be leveraged to improve image reconstruction quality. To efficiently enlarge the receptive field, we propose to augment the input image with circular-shifted replicas, which are then concatenated as additional image channels. The augmented input is fed into PCP-UNet with additional input channels (Fig. \ref{figCSCNN}). With a sufficient number of shifted replicas, this channel-shifting can enable having aglobal receptive field, while accepting arbitrary input sizes. Channel-shifting has a minor computational overhead equivalent to adding an additional convolutional layer with multiple channels. Channel-shifting has no additional memory consumption when the number of input channel is not greater than the maximum number of channels in the hidden layers. 

\subsection{\emph{In Vivo} CMR Dataset and Experiment Details}
We performed our experiments using the CMRxRecon dataset \cite{cmrxrecondata}, which contains fully sampled cine, aorta, tagging, and mapping (T1 and T2) CMR images from various views including short-axis, two, three, four-chambers, and long-axis views. Typically 5-10 slices and 12-25 cardiac phases were acquired. We split multi-slice acquisitions into single-slice 2D+time (2D+t) images and re-assembled the multiple slice after reconstruction. We used 150,480 2D+t images for training, and 4,158 images for validation. 

Each fully sampled image was retrospectively undersampled using uniform, Gaussian random and pseudo radial patterns at undersampling rates of 4, 8, 12, 16, 20 and 24. Coil sensitivity profiles were estimated using ESPIRiT \cite{espirit}. In reconstruction, complex images were mapped to two-channel real-valued 3D tensors. We compared the following network designs: \textbf{Fixed UNet}: Fixed amount of UI using a conventional UNet; \textbf{Adaptive UNet}: Adaptive unrolling using a conventional UNet; \textbf{Fixed PCP-UNet}: Fixed amount of UP using PCP-UNet; \textbf{Adaptive PCP-UNet}: Adaptive unrolling using PCP-UNet. In all tested methods, we used the conjugate gradient solver for data consistency. In all networks, we applied 2.5D convolutions, which perform 2D convolution in the image domain followed by 1D convolution along the temporal dimension.

\section{Results}

\begin{figure}
\includegraphics[width=\textwidth]{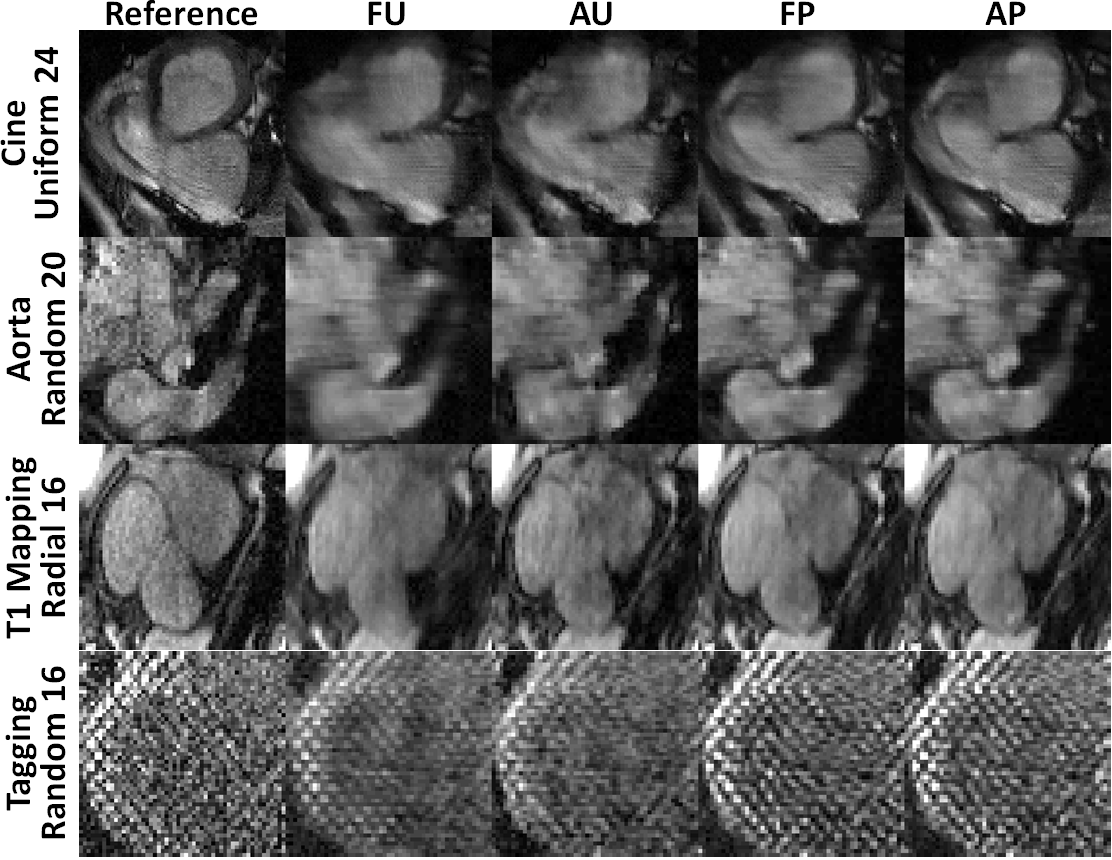}
\caption{Representative reconstructed image results for selected combinations of sampling patterns, acceleration rates, and contrasts, using all tested methods: \textbf{FU}: Fixed amount of UI with a UNet regularizer; \textbf{AU}: Adaptive amount of UI with a UNet regularizer; \textbf{FP}: Fixed amount of UI with the PCP-UNet regularizer; \textbf{AP}: Adaptive amount of UI with the PCP-UNet regularizer.} \label{FigGraphRes}
\end{figure}

For evaluation, we computed SSIM of single-slice 2D+t images with respect to fully sampled references, over a cropped central area covering 1/2 of $k_x$ and 2/3 of $k_y$ to remove background signals.

\begin{figure}
\includegraphics[width=\textwidth]{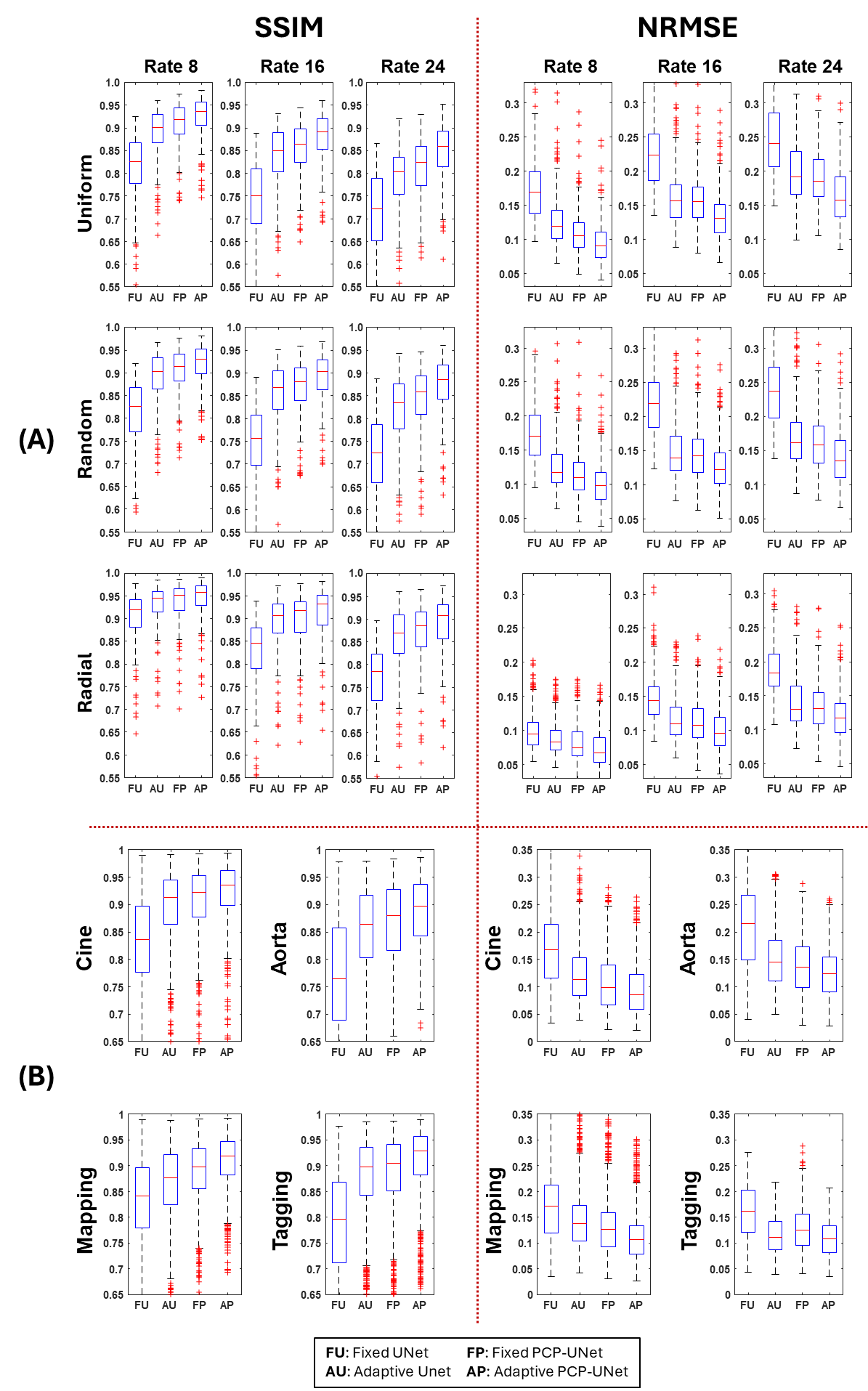}
\caption{SSIM and NMRSE statistics for each reconstruction method (Fixed UNet, Adaptive UNet, Fixed PCP-UNet and Adaptive PCP-UNet) with respect to: (A) the \emph{k}-space undersampling rate (8x, 16x, or 24x) and sampling pattern (uniform, random and radial), and (B) image contrast. AP generally outperforms (higher SSIM, lower NMRSE) the other methods.} \label{FigStatistics}
\end{figure}

\subsection{Experimental Results}
Fig. \ref{FigGraphRes} depicts representative reconstructed images for each reconstruction method, with selected combinations of sampling patterns, rates, and contrasts. In all cases, Fixed UNet shows visual blurring and aliasing artifacts. Adaptive UNet shows increased sharpness compared to Fixed UNet, but it still exhibits visible aliasing artifacts due to the high acceleration rate. Both Fixed and Adaptive PCP-UNet shows less visible aliasing. Adaptive PCP-UNet better recovers fine details compared to Fixed PCP-UNet.

Fig. \ref{FigStatistics} depicts the boxplots of SSIM and NMRSE for all tested reconstruction methods (Fixed UNet, Adaptive UNet, Fixed PCP-UNet and Adaptive PCP-UNet) and each undersampling pattern (uniform, random, radial) for a range of acceleration rates  (8x, 16x, and 24x), and for a range of image contrasts. Fixed UNet exhibits the lowest SSIM and PSNR among all tested methods. Adaptive UNet improves compared to Fixed UNet in terms of SSIM and NMRSE across different acceleration rates. Fixed PCP-UNet equipped with the addition of a pattern prompt further improved the SSIM and NMRSE across different acceleration rates compared to FU. Finally, Adaptive PCP-UNet combined both adaptive unrolling and a pattern prompt had the highest SSIM and lowest NMRSE among all tested methods. Paired t-tests show that the SSIM for each method is statistically different ($P = 0.05$) at all patterns, rates, and contrasts. 

\section{Conclusion and Discussion}
In this study we proposed a foundation model for general CMR reconstruction using adaptive unrolling and PCP-UNet equipped with a new channel-shift technique. We applied this to a variety of \emph{k}-space undersampling rates, \emph{k}-space sampling patterns, and image contrasts. Adaptive unrolling with a regular UNet significantly improved SSIM and reduced NMRSE for different acceleration rates. Employing PCP-UNet with adaptive unrolling further improved the image quality across different image contrasts. Our Adaptive PCP-UNet approach works very well for a wide range if CMR image contrast and acceleration rates. The proposed Adaptive PCP-UNet model has the potential to be fine-tuned for a specific reconstruction task. Transfer learning could be used to apply our method to other image anatomies, which will be studied in the future.

%


\end{document}